# Topic-based Integrator Matching for Pull Request

Zhifang Liao[1]    Yanbing Li[1]    Dayu He[1]    Jinsong Wu[2]    Yan Zhang[3]    Xiaoping Fan[4,*]

(1) Department of Software Engineering, Central South University, China
(2) Department of Electrical Engineering, Universidad de Chile, Santiago, Chile
(3) Department of Computer, Communication and Interactive System, School of Engineering and Built Environment, Glasgow Caledonian University, UK
(4) Hunan University of Finance and Economics, China

*Abstract*—Pull Request (PR) is the main method for code contributions from the external contributors in GitHub. PR review is an essential part of open source software developments to maintain the quality of software. Matching a new PR for an appropriate integrator will make the PR reviewing more effective. However, PR and integrator matching are now organized manually in GitHub. To make this process more efficient, we propose a Topic-based Integrator Matching Algorithm (TIMA) to predict highly relevant collaborators(the core developers) as the integrator to incoming PRs . TIMA takes full advantage of the textual semantics of PRs. To define the relationships between topics and collaborators, TIMA builds a relation matrix about topic and collaborators. According to the relevance between topics and collaborators, TIMA matches the suitable collaborators as the PR integrator.

## I. INTRODUCTION

With the development of technology, the human society has entered the big data age. More and more researchers has been attracted by the big data problem. GitHub, as the popular social coding community[1], also faces the big data problem. Currently, it has attracted 12 millions developers and 31 millions projects hosted on it. Pull Request (PR) is a primary method[2,3,4] for contributions from the external contributors. The recent research works have shown that the popular projects receive tens of PRs every day covering 60% of code commits from contributors. However, there is no automatic mechanism to assign the integrator to the PR. Matching an appropriate integrator of a new PR will make the PR review more effective, since it can reduce the latency between the actual review of the PR and the closure of the PR. To make this process more efficient, we propose a Topic-based Integrator Matching Algorithm (TIMA) to predict the highly relevant collaborators as integrator of incoming PRs. TIMA makes full use of the textual semantics of PRs, and matches the collaborators as the integrator of a new PR by the relationships between topics and collaborators.

The structure of this paper is as follows. Section 2 presents the previous research on PR. Section 3 describes the matching methodology--TIMA. In Sections 4 and 5, the experiment designs, results and validation are provided. And in Section 6, the potential problems are discussed in the research and a conclusion is drawn.

## II. RELETED WORK

The previous research works find that social medias benefit the PR review, like @-mention [4,5,6]. However, the developers will can't @ the suitable developers to review the PR when they are unfamiliar with each other. To solve this problem, Yue Yu[7] proposed a reviewer recommender combining information retrieval with social network analyzing to assign the suitable reviewers to new PRs.

Some other researchers have realized that Ensuring the quality of projects when merging PRs is also an essential problem [8,9,10] In order to ensure the quality of PR, Continuous Integration(CI)[11,12] is applied to the PR merging process in GitHub. When a new PR is committed, CI integrates the PR into a project automatically, and tests it. Bogdan Vasilescu et al.[13] proved that CI can ensure the quality of PRs via conducting a qualitative analysis study. Yue Yu et. al. [14] applied a linear regression algorithm to analyze the latency of PR process which used CI. Before using CI to improve the efficiency of PR review, the appropriate integrator should have been found. However, we have reviewed the currently existing research works about PR on GitHub, and have found that the research results about integrator and PR matching are very limited, which suggests great potentials for the efficiency improvement of PR review.

## III. METHODOLOGY

In GitHub, a PR is always reviewed by several collaborators of the project, they discuss the PR via comments. All the collaborators have the potential to become an integrator. In this paper, we treat each collaborator who reviewed a PR as integrator of the PR. Here, we extend the definition of integrator.

**Definition: If a collaborator reviewed a PR, then he is an integrator of the PR.**

We aim to match suitable collaborators as the integrator for a new PR. The representative existing works of reviewer recommendation were based on social networks. All these approaches start from constructing a developer social network, and then recommend reviewers through the social relationship of the submitter. In this

Supported by the Fundamental Research Funds for the Central Universities of Central South University

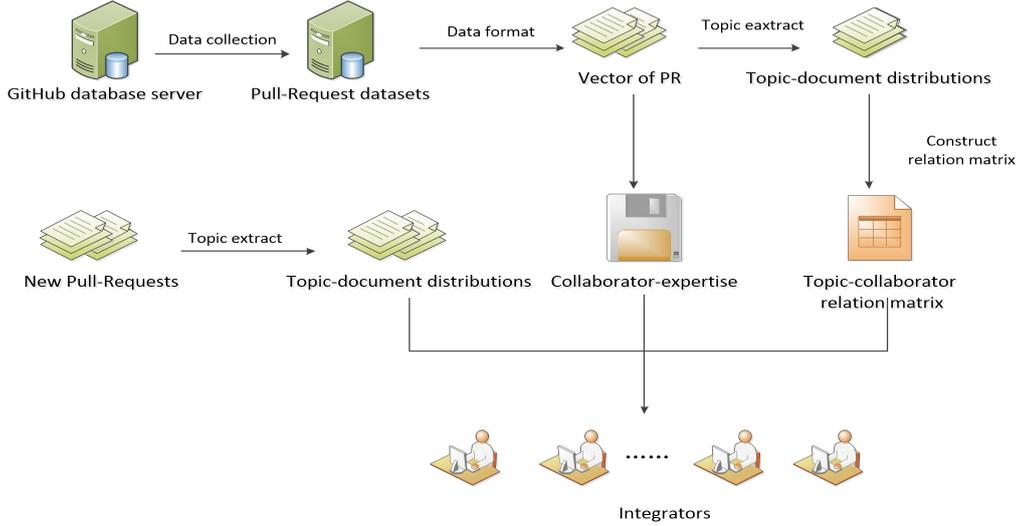

Fig.1 Overview of the proposed method TIMA. TIMA builds the collaborator-expertise and topic-collaborator relation matrix using the PR review history, and finds the most relevance collaborators as the integrator of the new PR.

paper, we propose a topic based method TIMA. TIMA treats each PR as a text document, and then constructs the relationship between collaborators and the topics extracted from PRs. Finally, TIMA matches the suitable collaborators as the integrator through the relationship between collaborators and topics. The overview of TIMA is shown in Fig.1. We describe more details in the following subsections.

*A. Vector of Pull Request*

Each PR can be characterized by its title, description and comments, and labeled with a set of collaborators who reviewed it. TIMA preprocesses the text of each PR via applying the required natural language text preprocessing steps used by any information retrieval technique. The preprocessing steps include tokenized, stop words removing, and stemming. Here, we adopt the Porter Stemmer as the stemmer algorithm. TIMA removes common English language stop words (such as: the, it, and on) to reduce noises by the stop-words list provided by Google[1]. In addition, TIMA stems the words(for example, 'fixing' becomes 'fix') in order to reduce the vocabulary size and reduce duplication due to the word form. Then, TIMA extracts topics from the text of PRs via applying topic generative model LDA [15]. Currently, the most popularity topic model is PLSA [16] and LDA. LDA is a upgraded version of PLSA which introduced the Dirichlet prior distribution. PLSA is more suite for the short text, while LDA is more suite for long text. Since the text of PR includes title, description and comments, it may more suite to be classified as long text. Hence, we adopt the LDA as the topic model. TIMA uses a vector to represent each PR as a weighted vector. The probability of each topic in a PR is a term.

*B. Relation Matrix Construction*

For each project, the corresponding *relation matrix* is constructed individually. In a given project, the structure of relationship between topics and collaborators may be a many-to-many model. Since each PR can be reviewed by more than one collaborators, and a collaborators also can review more than one PR. In each PR, the probability of each topic represents the importance of the topic, and each PR is labeled with a set of collaborators. Therefore, the probability of topics can reflect the relevance between topics and collaborators in each PR. However there are many PRs, so we need to calculate the topic-importance [17] of multi-document.

In a different document, the topic-importance should be different. Topic-importance is related to the length of the document. The longer a document is, the higher weight the topic-importance has in the document. Thus, the topic-importance should be calculated as follows,

$$import(t_i) = \frac{\sum_d^D N_d P(t_i|d)}{\sum_d^D N_d} \qquad (1)$$

where $t_i$ is the topic $i$; $D$ is the documents set; $N_d$ is the number of words in document $d$; $P(t_i|d)$ is topic-distribution for the document obtained by LDA model.

However, PR is always reviewed by several collaborators and each collaborator reviews a part of PRs. Therefore, we need do some changes of this equation, the relevance between topics and a collaborator should be represented by the topic-importance of the PRs which he reviewed. Thus, the relevance between collaborators and topics should be calculated as follows

$$R(r_i, t_i) = \frac{\sum_d^D \lambda_{di} N_d P(t_i|d)}{\sum_d^D \lambda_{di} N_d} \qquad (2)$$

where $r_i$ is the collaborator $i$; $\lambda_{di}$ is the control factor. If the PR is reviewed by the collaborator i, $\lambda_{di}$ is 1, otherwise $\lambda_{di}$ is 0; $d$ is the document; $D$ is the documents set.

However, the number of PRs reviewed by each collaborator is different. There may exist such a problem that the active collaborators are reviewed a lot of PRs, so that they will cover the relevance between topics and inactive collaborators. To classify the relevance between topics and collaborators, we need to normalize the relevance between them. Thus, the *relation matrix* should be calculated as

$$matrix(r_i, t_j) = \frac{R(r_i, t_j)}{\sum_k^K R(r_i, t_k)} \quad (3)$$

where $K$ is the number of topics extracted from PRs; $t_k$ is the topic $k$.

---
[1] https://code.google.com/p/stop-words/

*C. Collaborator-expertise*

Actually, the expertise of different collaborators may be different in a project. The more active the developer on a code, the higher his/her expertise on that piece of code is. In other words, the more PRs who reviewed, the higher his/her expertise is. Thus, the collaborator-expertise should be calculated as:

$$expertise(r_i) = \frac{\sum_d^D C(r_i, d)}{\sum_d^D \sum_r^R C(r, d)}, \quad (4)$$

where $D$ is the number of documents in the set; $R$ is integrator set, $C(r,d)$ is the number of integrator $r$ reviewed document $d$.

*D. Topic-distribution Calculation of New PR*

Based on the *relation matrix*, integrator matching for new PRs is divided into two steps. The first step is to calculate out the topic-distribution of new PRs. The second step is to match the suitable collaborators as the integrator for the new PRs according to the *relation matrix* and topic-distribution.

There are two ways to calculate the topic-distributions of new PRs. One is to put the new PRs into training set, and use LDA to extract the topic-distribution together, while the other one is to use the word-distribution of topics obtained by LDA model to calculate the topic-distribution of new PRs. Let's image such a scenes: We have a training set which include 100 PRs and 10 new PRs which unmatched integrator. If we adopt the first one, we need calculating 101*10 times(Since in practice, we need run the matching algorithm at the moment of new PR incoming); If we take the second one, we just need calculating 100+10 times. Obviously, the first one is more time-consuming than the second one. Although the second one receives a lower accuracy than the first one, it just requires to construct the *relation matrix* once. Compared with the first one, the second one is computationally more efficient. Considering the cost, TIMA adopts the second way to calculate the topic-distribution of new PRs.

The second way mainly uses the probability of words appearing in a topic to calculate the probability of topic appearing in a text. Thus, the topic-distribution of new PRs should be calculated as

$$P(t_i|d) = \frac{\sum_{w \in V} c(w,d) P(w|t_i)}{\sum_k^K \sum_{v \in V} c(v,d) P(v|t_k)} \quad (5)$$

where $c(w,d)$ is the number of word $w$ in document $d$; $P(w|t_i)$ is the word-distribution of topic $i$.

*E. Integrator Matching of New PR*

Since the topic-distribution of a new PR has been calculated, and the relationship between topics and collaborators also has been constructed, TIMA can find the closest collaborator according to the maximum entry of topic-distribution. If the topic-distribution has multiple maximum entry, it indicates the PR can be reviewed by multiple collaborators. As shown in **Algorithm 1**, TIMA finds the maximum entry of each PR first. Then, according to the maximum entry of each PR, TIMA calculates the matching score of each collaborator. Finally, TIMA selects the collaborator who get a highest matching score as the integrator. If there are more than one collaborators get the highest score, TIMA will match all of them to the PR as the candidates integrator.

| Algorithm 1 *Integrator Matching* |
|---|
| **Input:** Vector of New PR *V*, <br> Relation matrix *matrix(r<sub>i</sub>,t<sub>i</sub>)* |
| 1. *top_topics*=set() |
| 2. for *v* in *V*: |
| 3.    *max_topics*=arg max(*v*) |
| 4.    *top_topics*.add(*max_topics*) |
| 5. *candidates*=set() |
| 6. for *topics* in *top_topics*: |
| 7.    *candidate*=set() |
| 8.    for *k* in *topics*: |
| 9.       for *r* in *R*: |
| 10.         *s(r)*=*matrix(r,k)*expertise(r)* |
| 11.       *candidate*.add(arg max(*s(r)*)) |
| 12.    *candidates*.add(*candidate*) |
| **Output:** *candidates* |

IV. EXPERIMENTS

*A. Datasets*

We have chosen three popular open source projects hosted on GitHub as the datasets to train and test the method. The details of the datasets are shown in the Table 1. All of the datasets are collected by GitHub API. We collected all of the PRs which has closed by the end of 2016-08-01. Each PR consists of title, description, submitter, integrator, comments and reviewers. Since each collaborator is a potential integrator, the number of integrator is equal to the number of collaborators in the most situation. In this paper, we consider they are equal. In general, the number of developers can reflect the size of the project. In GitHub, the contributors equal to the developers. So, here we use the number of contributors to measure the size of the project. Based on the number of contributors, we have selected the project of various size(small, medium-base, large). The size of a project will defined as follows:

Small: The number of contributors less than 100.

Medium-base: The number of contributors more than 100, but less than 500.

Large: The number of contributor more than 500.

## B. Experiments Design

In this paper, we have to resolve such three questions:

**Q1**: What's the relationship between topics and collaborators? Is the relationship between topics and collaborators many-to-many or one-to-many or others?

**Q2**: How about the topic-distribution calculation method of new PR? Is the proposed topic-distribution calculation method executable?

**Q3**: What about the performance of TIMA? Is the TIMA effective for integrator matching of PRs?

Table 1    detail of datasets

| Project | # of PRs | # of collaborators | # of contributors |
|---|---|---|---|
| fastlane | 2868 | 15 | 602 |
| mopidy | 557 | 9 | 90 |
| coala | 939 | 30 | 213 |

For **Q1**, we will use thermodynamic diagram to visualize the results and explore the structure of it.

For **Q2**, we will use Jensen-Shannon divergence[18] to measure the divergence of topic-distribution between LDA and TIMA. It is a symmetrical version of Kullback-Leibler divergence, the divergence is calculated as

$$D_{js}(p,q) = \frac{1}{2}(D_{kl}(p, \frac{p+q}{2}) + D_{kl}(q, \frac{p+q}{2})) \quad (6)$$

For **Q3**, we will use precision and recall to verify the performance of TIMA. Since there no a fixed value of number of topics extracted by LDA, here we'll extracted different number of topics from PRs, and compare the performance between them.

In this paper, we adopt the JGibbsLDA, a implementation that uses Gibbs sampling. And the hyper-parameter alpha, beta is using the default value, the value of iteration is 1000, the number of topics is 15. Since LDA is a probabilistic topic model, it may return different results if executed multiple times. But the distribution of the topic-distributions always be similar in general, and we just need the relationship between topics and collaborators. We don't care about what the topic is. For example, we execute the LDA two times, and get two topic-distributions. May be the implied theme of the topics is changed, topic1 int the first result may be similar with topic5 in the second result. We always can find the correspond topic between the two results. Indeed, what we want to get is the relationship between collaborators and the implied theme of the topics. So, the problem LDA may return different results if executes multiple times never influence the matching results.

## C. Evaluation Method

In this paper, we evaluate the performances of TIMA over each PR by precision and recall which are widely used as standard metrics in previous works. In GitHub, a PR is always reviewed by several collaborators(the core team members), each one of them is a potential integrator of the PR. If they reviewed a PR, here we call them core reviewer of the PR. Since they have the potential to become integrator, we treat them as the candidates of integrator. If the matched integrator of the PR is included by the core reviewers, we think the matching result is right. The formula of our metrics are list below

$$\text{Pr}ecision = \frac{|core\_reviwers \cap match\_set|}{|match\_set|} \quad (7)$$

$$\text{Re}call = \frac{|core\_reviwers \cap match\_set|}{|core\_reviwers|} \quad (8)$$

where *core_reviewers* is the set of collaborators who reviewed the PR; The *match_set* is the set of integrator of the PR matched by TIMA.

## V. RESULTS

### A. The Structure of Relation Matrix

For **Q1**, we have constructed a *relation matrix* about collaborators and topics by TIMA, and visualized the results by thermodynamic diagram. As shown in Fig.2, it's a *relation matrix* of project *fastlane*, the grid cell intensity is mapped to the relevance between collaborators and topics. The stronger the intensity, the higher the relevance between collaborator and topic. From Fig.2, we can find that some topics can be covered by most collaborators, while some topics can be addressed only by certain collaborators. In this paper, the topics are always mapped to features of a project, which means that the number of collaborators related to a topic reflects the popularity of the feature which the topic is mapped to. The more collaborators related to the topic, the more popularity the topic is. For example, topic7, topic16, topic17, and topic18 are more popularity than the others, since they are related to a lot of collaborators. From the thermodynamic diagram of *relation matrix*, we not only can obtain the relationship between collaborators and topics, and also can analyze the popularity of each topic.

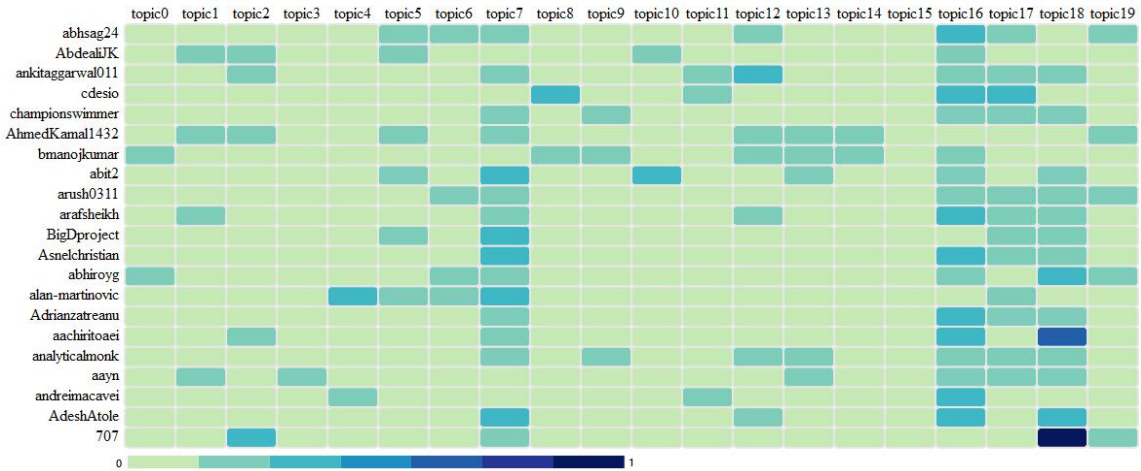

Fig.2 Thermodynamic diagram of *relation matrix* in *fastlane*. Grid cell intensity is mapped to the relevance between collaborators and topics. The stronger the intensity, the higher the relevance between collaborator and topic. The results show some topics can be covered by most collaborators, while some topics can be addressed only by certain collaborators.

*B. Topic-distributions*

For **Q2**, we have calculated the topic-distributions by the TIMA, and compared with the topic-distributions calculated by LDA. We have randomly selected 10 topic-distributions calculated by LDA and TIMA respectively. To clarify the difference between them, we have visualized the distributions, and compared them. Fig.3 shows the results, where vertical axis denotes the documents, and the horizontal axis denotes the topic-distributions. As shown in Fig.3, the maximum topic in each document is consistent in the most situations, and the difference of the topic-distributions with low probability is significant. On the one hand, they own a low probability, which means that they are not important in a document. On the other hand, we match integrator of a PR through the maximum topic. Thus, the difference of topics with low probability does not affect the results. Moreover, we have selected 1000 documents to calculate the topic-distributions by LDA and TIMA, respectively, and used Jensen-Shannon divergence to evaluate the divergence between them. The ranges of Jensen-Shannon divergence is [0,1]. The larger the value is, the more significant the divergence between two distributions is. We have calculated the Jensen-Shannon divergence of 1000 documents, and the average divergence is just 0.050. The average divergence close to 0, which means that the two distributions are very similar. Thus, it is feasible to use TIMA to calculate the topic-distributions of new PRs.

*C. Performance*

For **Q3**, we have conducted multiple sets of comparative experiments. We have selected project *fastlane*, *coala* and *mopidy* as the test datasets, and extracted topics with K(number of topics)=10, 15, 20, 25, 30 and 40, respectively. The results are shown in Fig.4. As shown in Fig.4(a), the precision of integrator matching is stable. Whatever the value of K is, the fluctuation of the precision is very small. The minimum precision is greater than 60%, close to 70%, while the maximum precision is close to 90%. From Fig.4(b), we can find that the fluctuation of number of topics to the recall is small too. But we also can see that the recall get a low score, the average recall is around 50%, and the maximum recall also around 60%. Since TIMA just selects the collaborators who get the maximum matching score as the integrator, but the score of different collaborator is always different. So, the recall always get a low score.

External contributors are not always familiar with the core developers. Thus, when they submit new PRs, they do not know who is the suitable integrator for his or her PR. They cannot use the "@-mention" to "@" the suitable reviewers. In that situation, the core developers should check the PRs one by one, and assign the suitable integrator for PRs, which is a heavy workload for the core developers. The delay of processing may be very long. Although TIMA just can provide a precision of 70%, there still exists 70% possibility that a PR can get suitable integrator. That means TIMA can match the suitable integrator to the PR in the most situation. Meanwhile, the core developers who receive the error message, they can "@" the suitable integrator to process this PR (since they are familiar with each other). Thus, the efficiency of PR reviewing is improved.

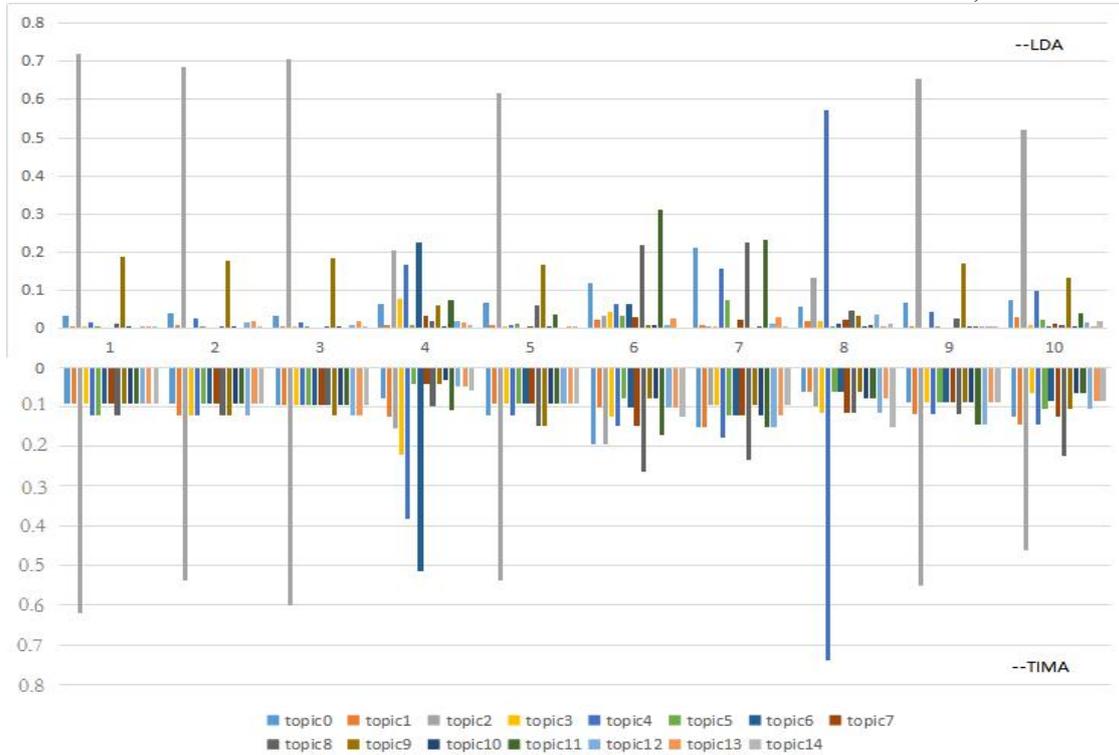

Fig.3 Topic-distributions comparison between JGibbsLDA and TIMA. 15 topics are extracted by each method. The maximum topic of the distribution is consistent in each document mostly, and on average the Jensen-Shannon divergence between the two distribution is just 0.05 which means they are very similar.

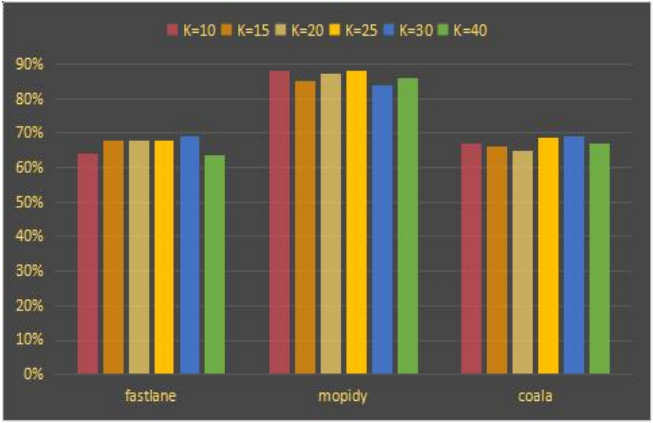

(a) Precision

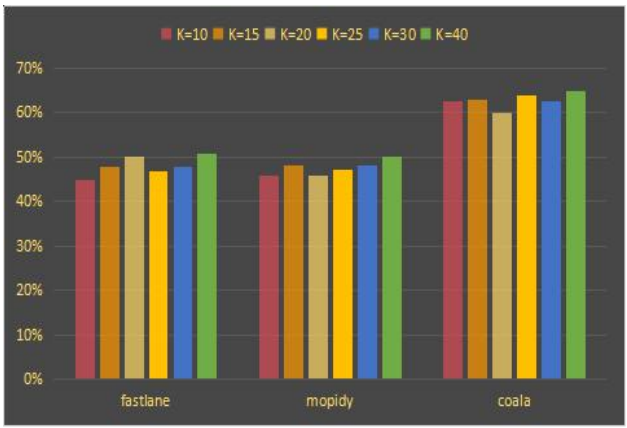

(b) Recall

Fig.4 Precision and recall of three project(*fastlane, mopidy, coala*). The number of topics extracted from PRs is 10, 15, 20, 25, 30 and 40, respectively. With the changing of number of topics, the fluctuation of the precision and recall is very small. On average, the precision is greater than 70% and the recall is greater than 50%.

## VI. DISCUSSION & CONCLUSIONS

### A. Discussion

*Limitation of TIMA algorithm:* The performance of TIMA depends on the history of the project. If there no enough closed PR histories, TIMA will get a poor performance, especially when meet a new project. Since TIMA only selects the collaborators who get the highest score as the candidates of integrator, the TIMA always get a low score of recall.

*Heavy workload for some very active collaborators:* From the observation, we found that some collaborators viewed a lot of PRs. Since TIMA matches integrator to a PR using the history of PRs, it is possible the very active collaborators would be frequently matched. Consequently, they would be burdened with a huge number of assigned PRs. Thus, considering workload balancing would reduce tasks of these potential integrator.

### B. Conclusions

In this paper, we have proposed an automatic integrator matching algorithm for PRs based on textual information contained in each PR. We take full use of the textual semantic of PRs, build the relationship between topics and collaborators. According to the relationship between topics and collaborators, we match each PR of suitable collaborators as integrator. We have tried to improve the efficiency of PR reviewing via matching suitable integrator of new PRs. Based on three open-source software project(*fastlane, coala* and *mopidy*), we have explored the *relation matrix*, topic-distributions and the precision. After evaluating the results, we have concluded that:

*1)* .Some topics can be covered by most collaborators, while some topics can be addressed only by certain collaborators. .

*2)* .The divergence of topic-distributions between LDA and TIMA is very small, just 0.050. They are very similar.

*3)* .On average, TIMA can reach a precision of 70%, while the minimum precision is greater than 60%. And the number of topics does not affect the precision.

In the future, we will continue to optimize our algorithm, employ TIMA to implement an integrator matching system, applying it to the GitHub.

ACKNOWLEDGMENT

The works described in this paper supported by Supported by the Fundamental Research Funds for the Central Universities of Central South University.